\documentclass[twocolumn,amsmath,amssymb,prl]{revtex4}
\usepackage{graphicx} \usepackage{bm}

\begin{document}

\title{Ultrashort pulse propagation and  the Anderson localization}

\author{S. Gentilini$^{1,2}$,   A. Fratalocchi$^{3,1}$,
L. Angelani$^{4}$ G. Ruocco$^{1,2}$, C. Conti$^{2}$}

\address{$^1$Dept. of Physics, University Sapienza, Piazzale Aldo Moro 5, I-00185,   Rome,  Italy \\
$^2$Research Center Soft  INFM-CNR, c/o Dep. Physics, University  Sapienza, Piazzale Aldo Moro 5, I-00185, Rome, Italy\\ 
$^3$Research  Center ``Enrico Fermi'', Via Panisperna 89/A, I-00184, Rome, Italy \\
$^4$Research Center SMC  INFM-CNR, c/o Dep. Physics, University  Sapienza, Piazzale Aldo Moro 5, I-00185,  Rome, Italy}

\begin{abstract}
We investigate the dynamics of a $10$~fs light pulse propagating in a random medium by 
the direct solution of the 3D Maxwell equations.
Our approach  employs molecular dynamics to generate a distribution of spherical scatterers and
a  parallel finite-difference time-domain  code for the  vectorial wave propagation.
We calculate the disorder-averaged energy velocity and the decay time of the transmitted pulse Versus the localization length
for an increasing refractive index. 
\end{abstract}



\maketitle

As  originally discussed by  Anderson 
\cite{Anderson57}, and more recently in several articles
\cite{john84,Kaveh,Ishimaru84,Albada85,Maret,Wiersma97,Johnson03,Pinheiro04,Lubatsch05,
Wiersma07,Skipetrov07,toninelli:08,Conti08},
including Bose-Einstein condensation \cite{Billy08,Inguscio08}, elastic networks \cite{Hu08},
and optical beams \cite{Segev07},
three-dimensional (3D) wave localization may occur in the presence of structural randomness.
An interesting issue is the role of localized states in the propagation of ultrashort laser pulses in 
random media  \cite{Calba:08}, eventually including nonlinear effects \cite{Conti07}.

In this Letter we report on ab-initio computational results of ultra-short pulses in random 
media with increasing refractive index.
Our approach combines Molecular Dynamics (MD)  and parallel Finite-Difference  Time-Domain (FDTD) codes: the  former to
provide realistic 3D distributions  of scatterers with
quenched disorder,  the latter to exactly solve the vectorial  Maxwell equations \cite{Conti07}. 
With such  MD-FDTD  technique, we study the  propagation of classical waves in
3D disordered dielectrics  for  a varying  scattering
strength, as obtained  by changing the scatterer refractive
index  $n$. As $n$ grows, the effective optical path increases and localization  occurs.   
In what follows, we characterize the degree of localization of the electromagnetic (EM) field  
by the inverse participation ratio $\xi$ of the 3D energy profile and relate it
to the transmission delay (expressed in terms of the energy velocity $v_e$)
and to the decay constant of the trailing edge of the transmitted pulse (expressed in terms of an effective 
diffusion  coefficient $D$). 
As the strong-localization regime is attained, $D$ and $v_E$ decrease; 
in addition the spectrum of the transmitted pulse displays several narrow peaks.
By relating  $v_E$ to $\xi$, we  observe a signature  of a
transition,  above which  localized  states  are present.
\begin{figure}[htb]
\centerline{
\includegraphics[width=8.3cm]{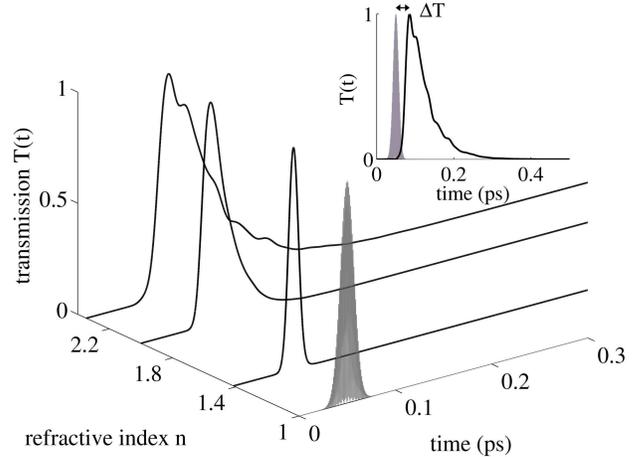}}
\caption{  (Color online).  Transmitted  pulse $T(t)$  in
linear scale for increasing refractive index $n$; the filled region is the adopted input
pulse. $T(t)$ is scaled to unitary peak value for any $n$.
The inset shows the time delay $\Delta T$ used to evaluate the energy propagation velocity $v_E$.
\label{transmittedpulse}} 
\end{figure}

Our  sample is a  distribution of  1000 spherical  scatterers obtained
by MD simulations. 
Particle dimensions  are chosen in order to match
typical    samples     used    in    experiments,     as    e.g.    in
\cite{Vellekoop05,Storzer06}.  We use a $50/50$ mixture
of particle diameters $310$~nm and $248$~nm interacting with a generalized Lennard-Jones potential \cite{Conti07};
at a filling fraction $\phi\cong0.6$, 
this results in a largely disordered and tightly packed particle distribution
in a cube with edge $L=2.9~\mu$m. The  refractive index of the scatterers
$n$ is chosen in the experimentally accessible range between $1.2$ and $2.8$. 
Several MD  runs  furnish different configurations of the disorder.  

For each  realization, we  solve the
Maxwell  equations  by a parallel FDTD algorithm \cite{TafloveBook}  and  study the  transmission  of  a
Gaussian  TEM$_{00}$ linearly  $y-$polarized input  pulse,  with waist
$w_0=1~\mu$m, impinging on the $xy$ face of the cube at normal incidence. The input pulse temporal profile is also Gaussian, with
duration $t_0=10$~fs and  carrier wavelength $\lambda_0=532$~nm. 
Numerical  results  have  been  averaged  over  five  MD
configurations of the $1000$ colloidal particles.
\begin{figure}[htb]
\centerline{ 
\includegraphics[width=8.0cm]{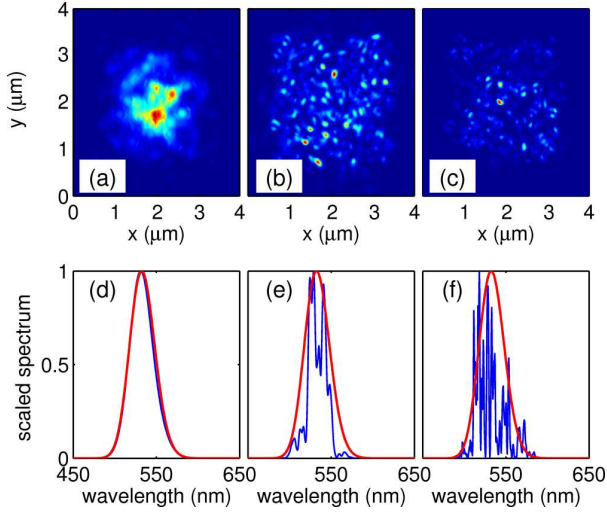}}
\caption{(Color  online). 
(a-c) spatial distribution of the energy at the output face of the sample (at $t=0.5$~ps for a CW excitation at $\lambda=532$~nm). 
(d-f) spectrum of  the electric field $E_y$ (thin line) calculated for $n=1.4$,  $2.2$ and $n=2.8$; 
the thick line is the spectrum of the incident pulsed beam. 
\label{figspectrum}}
\end{figure}
For  each  set  of  MD-FDTD   simulations  we  collect:  i)  the  total
transmission $T(t)$ by integrating  the $z-$component of the Poynting
vector with  respect to the  transverse ($x,y$) output plane,  ii) the
$y-$component   of  the   output   electric  field   $E_y$,  iii) the corresponding
spectrum and iv) the EM energy  density   $\mathcal{E}$.  
In addition, we calculate   the  distribution  of  decay  times
$g(\tau)$:
\begin{equation}
  \label{eq:0}
  T(t)=\int_0^{\infty} e^{-t/\tau}g(\tau) d\tau,
\end{equation}
by a best-fit with a  superposition of exponentially  decaying functions.  
The mean value $\bar\tau$ for the decay-time is then used to calculate 
the ``effective'' light diffusion constant as $D=L^2/\pi^2\bar\tau$
(in the broadband short-sample regime here considered the diffusion approximation is not expected to be
strictly valid).The  energy propagation 
velocity $v_E$ is calculated by determining  the time $\Delta T$ spent by the pulse peak (from the Poynting vector) 
to travel from the input to the  output face of  the sample (see inset in Fig. \ref{transmittedpulse}),
letting $v_E=L/\Delta T$, and averaging over disorder realization. 
\\To characterize the  localization length, we  calculate the inverse participation
ratio $\xi$:
\begin{equation}
\label{partecipation}
  \xi=\sqrt[3]{\frac{(\int_V\mathcal{\mathcal{E}}^2(x,y,z)dV)^2}{\int_V\mathcal{\mathcal{E}}^4(x,y,z)dV}},
\end{equation}
being $V=L^3$ the sample volume. $\xi$ is such that if the energy profile decays as an exponential
with decay constant $l_{\xi}$, it is $\xi=2l_\xi$; hence it directly measures the spatial extension
of $\mathcal{E}(x,y,z)$. $\xi$ is calculated by using a continuous wave (CW)
beam at $\lambda=532$~nm, to avoid the simultaneous excitation
of several modes.
\begin{figure}[htb]
\centerline{
\includegraphics[width=8.3cm]{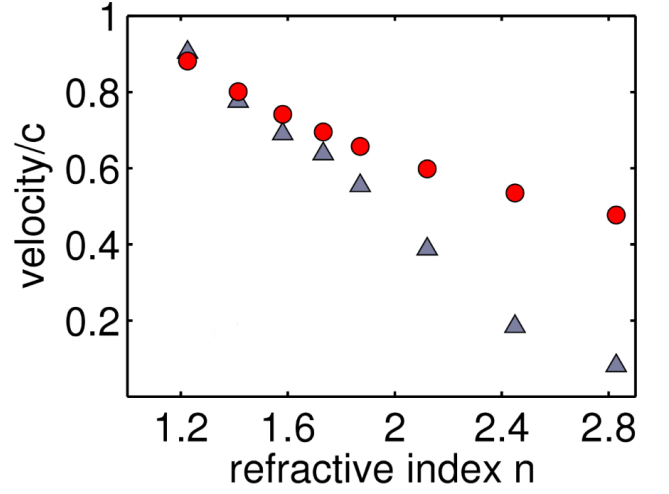}}
\caption{(Color  online). Phase  $v_p$ (circles) and  energy  $v_E$
(triangles) velocity versus $n$.
\label{ve}}
\end{figure} 

Figure  \ref{transmittedpulse} shows the  input (filled region)
and the transmitted pulses $T(t)$ (solid lines), for increasing values
of $n$; the $T(t)$ tail gets longer due to the reduced light diffusivity $D$, while the transmitted pulse slows down
(see figure \ref{ve} below).
As shown in  Fig. \ref{figspectrum}a-c, 
spatial distribution  of the energy $\mathcal{E}$ radically
changes from extended $(n=1.4)$ to localized states $(n=2.2$ and
$n=2.8)$.  
Correspondingly,  the spectrum  of  $E_y$ splits into multiple modes 
(Fig. \ref{figspectrum}d-f), which implies longer lifetimes
for the involved EM resonances and     a     dynamic slowing     down (Fig.  \ref{transmittedpulse}).  

In Fig.  \ref{ve} we compare the trend of $v_E$ (triangles)
versus $n$ with that of the corresponding phase   velocity   $v_p$  (circles),
which is calculated as $v_p\simeq c/\bar{n}$ [being $\bar{n}=\phi~n+(1-\phi)$, the mean
refractive index of the colloidal spheres dispersed in air]. 
The increase of the degree of localization is accompanied by the concurrent  swelling of the
discrepancy  between $v_E$  and  $v_p$,  which
becomes appreciable for $n$ greater than a critical value $n_c\cong1.8$.
Previous experimental investigations have reported  significant deviations from an exponential 
transmission for an average index $\bar n=1.55$ \cite{Storzer06}; in our case we have as the critical value
for the localization $\bar n=1.48$. However further work is required for a strict 
quantitative comparison with experiments.
\begin{figure}[htb]
\centerline{
\includegraphics[width=8.3cm]{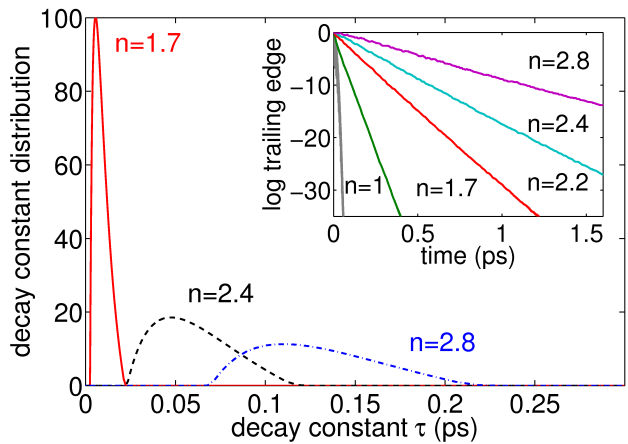}}
\caption{(Color online).  Distribution of decay-times  of the trailing
edge  of the  transmission  $T$ for various $n$.
The inset shows the log-scale plot of normalized transmitted signals $T(t)$ (trailing edge).
\label{times}}
\end{figure}

Figure \ref{times}  shows the  decay constant distribution $g(\tau)$ of trailing edge of the transmission
$T(t)$, as  calculated from Eq. (\ref{eq:0}).  
The inset of  Fig. \ref{times} shows the trailing edge of the scaled and 
temporally shifted transmitted pulse in logarithmic scale;  the reported linear trends apparently 
indicate the absence  of any sensible deviation from a single exponential,
which however becomes evident from the spreading of $g(\tau)$.
Note that at the localization, the width of $g(\tau)$ is comparable
to its mean value $\bar\tau$.
\begin{figure}[htb]
\centerline{
\includegraphics[width=8.3cm]{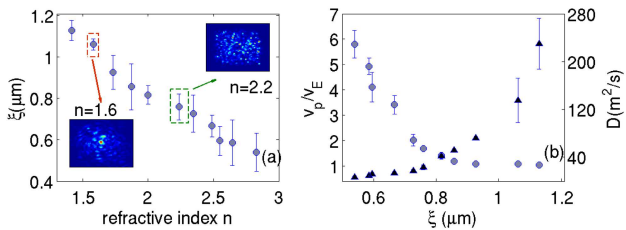}}
\caption{(Color   online).(a) Localization  length   $\xi$  versus
refractive index $n$; the error bars and average values as calculated over $5$ realizations of the disorder.
The two insets show the snapshot of the energy profile at the output face of the sample for two values of $n$.
(b)Parametric  plot (versus $n$) of the ratio $v_p/v_e$ (left scale) and
of $D$ (right scale) versus $\xi$.
\label{localization}}
\end{figure} 

Figure  \ref{localization}a shows the participation  ratio  $\xi$ for
different values of the refractive  index $n$.
As expected the EM resonances become more localized as 
the optical path in each scatterer increases; however since
$\xi$ is a volume averaged quantity, it does not
display a discontinuous trend versus $n$.
In Fig. \ref{localization}b  we draw  the ratio  between phase  and
energy velocity $v_p/v_E$  and the dynamic diffusion $D$ versus $\xi$.
This analysis yields  a crossover  at $\xi(n_c)$,
where  the  discrepancy between  $v_E$  and  $v_p$ is evident. 
Beyond the threshold the reduction of the localization length is accompanied
by the slowing down of the pulse and a simple
one-to-one relation between $\xi$ and $v_E$ is evident.
It is important to stress that $D$ is expected to vanish at the 
Anderson transition for an infinitely extended structure;
here we find that, for finite size systems, $D$ (and the energy velocity $v_E$) directly measures 
the degree of spatial localization.

In conclusion we reported on what we believe to be the first
time-resolved analysis of ultra-short light pulses in 
3D disordered samples.
Our approach combines molecular dynamics and finite-difference
time-domain  codes,  thus providing  a  realistic
distribution of the scatterers and an exact theory of wave propagation. 
The distribution of the decay-time is shown to largely spread
at the Anderson transition.
The plot of the ratio between the energy and the phase velocity
versus the localization length displays a critical character;
the more localized are the excited EM resonances, 
the slower is the input pulse propagation.
The diffusion constant is not vanishing at the localization transition and it is a direct measure of the
spatial extension of the EM field.
These findings are expected  to stimulate novel theoretical works  and experiments in
the  large community  dealing  with energy  propagation in the presence of disorder, ranging from optics to quantum systems.
\\{\it Acknowledgments. ---}
The research  leading to these  results has received funding  from the
European  Research  Council  under  the European  Community's  Seventh
Framework Program  (FP/2007-2013)/ERC grant agreement  n. 201766.  We
acknowledge support  to INFM-CINECA and CASPUR for  the initiative for
parallel computing.

\bibliographystyle{ol}

\end{document}